# Plasmon-enhanced electron-phonon coupling in Dirac surface states of the thin-film topological insulator $Bi_2Se_3$

Yuri D. Glinka,[1,2]* Sercan Babakiray,[1] and David Lederman[1]

[1]Department of Physics and Astronomy, West Virginia University, Morgantown, WV 26506-6315, USA
[2]Institute of Physics, National Academy of Sciences of Ukraine, Kiev 03028, Ukraine

Raman measurements of a Fano-type surface phonon mode associated with Dirac surface states (SS) in $Bi_2Se_3$ topological insulator thin films allowed an unambiguous determination of the electron-phonon coupling strength in Dirac SS as a function of film thickness ranging from 2 to 40 nm. A non-monotonic enhancement of the electron-phonon coupling strength with maximum for the 8 - 10 nm thick films was observed. The non-monotonicity is suggested to originate from plasmon-phonon coupling which enhances electron-phonon coupling when free carrier density in Dirac SS increases with decreasing film thickness and becomes suppressed for thinnest films when anharmonic coupling between in-plane and out-of-plane phonon modes occurs. The observed about four-fold enhancement of electron-phonon coupling in Dirac SS of the 8 – 10 nm thick $Bi_2Se_3$ films with respect to the bulk samples may provide new insights into the origin of superconductivity in this-type materials and their applications.

## I. INTRODUCTION

The measurement and control of the electron-phonon coupling strength (usually denoted as $\lambda$ – the dimensionless electron-phonon coupling constant) in topological insulators (TIs), such as $Bi_2Se_3$,[1-3] is one of the most important problems of these unique materials since their discovery[4,5] for several reasons. From a fundamental physics point of view, TIs offer an opportunity to study the effect of the renormalization of phonon energies between the insulator-type bulk (3D) states (the bandgap of $Bi_2Se_3$ is ~0.3 eV) and the metal-type two-dimensional (2D) Dirac surface states (SS) on photoexcited carrier relaxation dynamics.[3,6-8] Because the electronic and ionic subsystems in 3D states and 2D Dirac SS interact with each other through different mechanisms [electron-polar-phonon (Fröhlich) coupling and deformation potential/thermoelastic scattering, respectively], the carrier relaxation rate strongly depends on where photoexcited carriers are located during their energy relaxation. Switching between these two relaxation channels, which can be achieved by varying the thickness of $Bi_2Se_3$ films,[6-9] allows controlling the photoexcited carrier relaxation rate and hence the speed of novel optoelectronic devices based on the thin-film TIs.

On the other hand, the electron-phonon coupling has also been suggested to be one of the key factors determining the superconducting states in doped TIs, such as $Cu_xBi_2Se_3$,[10-12] in conventional electron-phonon superconductors[13-16] and in unconventional superconductors, such as cuprates, $Bi_2Se_3$ under pressure, and FeSe.[17-19] Strong electron-phonon coupling in these systems immediately points to the possibility of superconductivity because, according to the BCS theory, electrons are expected to become paired by (attractive) exchange of virtual phonons, which prevails over the electron-electron repulsion if the electron-phonon coupling is sufficiently strong.

In this paper, we present experimental results on Raman scattering studies of epitaxial thin films of the TI $Bi_2Se_3$ ranging in thickness from 2 to 40 nm. The film thickness and roughness is precisely determined using x-ray reflectivity (XRR) measurements. The thickness-dependent strength of electron-phonon coupling in Dirac SS is determined from the Fano-type resonance appearing for the surface phonon mode of the topmost hexagonally arranged continuous network of Se-Se bonds (H-mode) associated with Dirac SS. The observed non-monotonic enhancement of the electron-phonon coupling strength with decreasing film thickness, with a maximum at 8 - 10 nm, is suggested to originate from plasmon-phonon coupling which enhances electron-phonon coupling when free carrier density in Dirac SS increases with decreasing film thickness and becomes suppressed for thinnest films when anharmonic coupling between in-plane and out-of-plane phonon modes occurs. The plasmon-enhanced electron-phonon coupling strength in Dirac SS of the 8 - 10 nm thick films is approximately four times greater than the known value for bulk $Bi_2Se_3$ ($\lambda = 0.08$).[2] This observation may provide new insights into the origin of superconductivity in this type of materials in particular and layered structures in general.

## II. EXPERIMENTAL DETAILS

Experiments were performed on $Bi_2Se_3$ thin-film samples that were 2, 4, 5, 6, 8, 10, 12, 15, 20, 25, 30, 35, and 40 quintuple layers (QLs) thick (QL ~ 0.954 nm). The films were grown on 0.5 mm $Al_2O_3$(0001) substrates by molecular beam epitaxy and capped in-situ with a 10 nm thick $MgF_2$ layer to protect against oxidation as described previously.[6,20] The thin film thickness was determined from x-ray reflectivity (XRR) measurements analyzed using the GenX software.[21] Reflection high energy electron diffraction (RHEED) and x-ray diffraction (XRD) were used to analyze the structure of the $Bi_2Se_3$ films to determine that all films are epitaxial in the plane and highly crystalline out of the plane of the film.[20] Transport measurements revealed that all films have $n$-doping level in the range of $0.5 - 3.5 \times 10^{19}$ cm$^{-3}$, which is typical for as-grown $Bi_2Se_3$ films and bulk single crystals.[9]

Raman scattering measurements were performed using a Renishaw inVia Raman Spectrometer equipped with 785 nm solid-state laser. The spectra were acquired in a backscattering geometry with 1 cm$^{-1}$ steps and with laser power of 30 mW. A microscope of the spectrometer with a 50× objective lens was used to focus the laser light on the sample surface to a spot size of ~1 μm in diameter. The corresponding power density was $I_L = 3.8$ MW/cm$^2$. The $z(x,x)\bar{z}$ geometry was used, where $z$-axis is directed along a film normal and $x$-axis points to the light polarization.

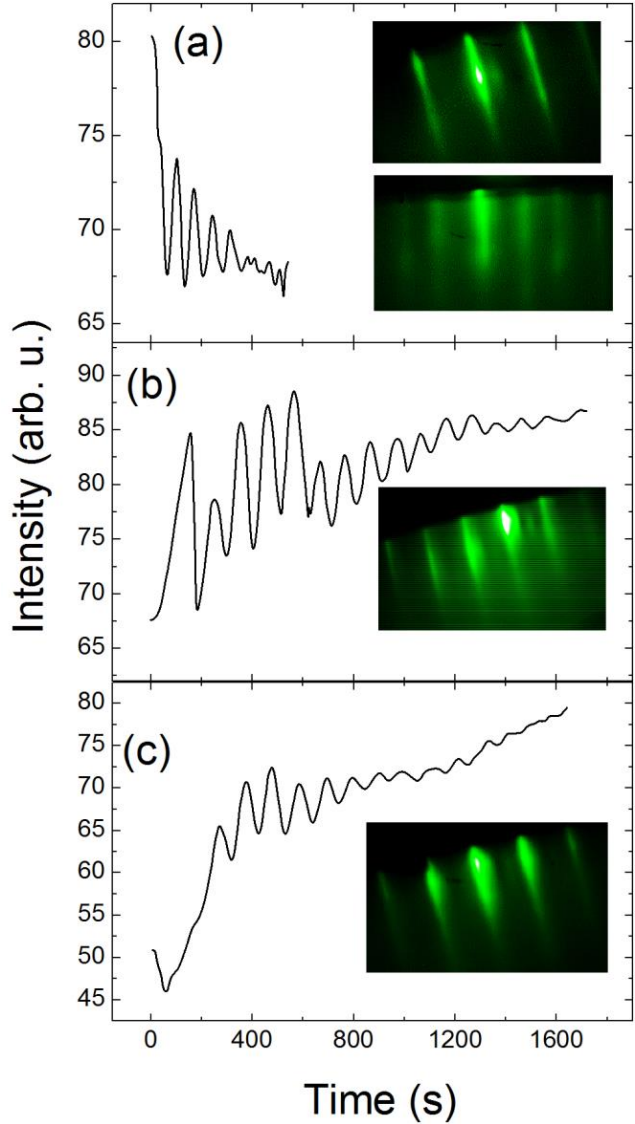

FIG. 1. RHEED intensity of the specular diffraction spot as a function of time for the (a) 10, (b) 20, and (c) 40 QL samples. The RHEED oscillations for the 40 QL sample were not discernible after 1600 s and the data acquisition was then stopped. The data were acquired for the high temperature growth after the initial deposition of 3 QL. Insets show the RHEED patterns after growth. The top inset corresponds to the RHEED pattern of the 4 QL sample which was grown at low temperature and then annealed, and therefore no RHEED oscillation pattern was recorded.

## III. EXPERIMENTAL RESULTS AND DISCUSSION

### A. Structural characteristics and film thickness determination

Typical RHEED data are shown in Fig. 1. RHEED oscillations were observed for samples with nominal thicknesses of approximately less than 20 QL, indicating layer-by-layer growth for those samples, as discussed previously.[20] Because the samples were grown using a two-step process, with the first 3 QL grown at low temperature (140 °C) prior to growth of the rest of the sample at a higher temperature (275 °C),[6] all samples with nominal thicknesses less than 4 QL were grown at low temperature and then annealed. All of the samples had streaky RHEED patterns indicating epitaxial growth as shown in the insets of Fig. 1.

Figure 2 shows the x-ray reflectivity intensity as a function of momentum transfer vector $q$ perpendicular to the surface for samples with nominal thicknesses of 4, 10, 20, and 40 QL. Similar results were obtained for the other samples used in this study. For samples less than 12 QL thick, the finite-size fringes for the (003) Bragg peak overlapped with the low $q$ oscillations, and therefore it was necessary to use a multilayer approach to fit the data out to the position of the Bragg peak. For these samples, the model consisted of Bi$_2$Se$_3$ QLs separated by a spacer layer representing the van der Waals coupling between QLs. For all samples, an interface layer at the top and/or bottom consisting of the average electron density of the substrate and Bi$_2$Se$_3$ (for the bottom) or of MgF$_2$ and Bi$_2$Se$_3$ (for the top) was included in order to simulate larger interface disorder parameters. The scattering length density $\rho_{xr}$ and its derivative $d\rho_{xr}/dz$ as a function of depth $z$, derived from fitting the layer thickness, atomic density, and interface roughness parameters, were used to determine the actual layer thicknesses and interface roughnesses, as described in detail previously in Ref. 22. The parameters obtained from this procedure are shown in Fig. 2. For the 4 and 10 QL sample data shown in Figs. 2(a) and (b), the number of maxima in $\rho_{xr}$ in the Bi$_2$Se$_3$ region is equivalent to the number of QLs of the sample. Using the 9.540 Å for the lattice parameter corresponding to a single QL (1/3 of the $c$-axis in the hexagonal structure, 28.620 Å) at room temperature,[23] the actual number of QLs for 4, 10, 20, and 40 QL samples were 4.6, 10.4, 17.8, and 43.5 QLs, respectively, with an uncertainty of approximately 5%. The roughness parameter at the Bi$_2$Se$_3$ interface increased progressively as the samples became thicker, which is consistent with the decreased amplitude in RHEED oscillations as the sample were grown thicker, which is probably due to an increase in disorder due to lateral regions with different number of QLs.

The structure analyzed via XRD with $q$ parallel to the growth direction is shown in Fig. 3. For all samples, the (003n) peaks (n is an integer) were clearly observed, indicating a [001] growth direction. An analysis of the (006) peak near $q = 1.32$ Å$^{-1}$, using the equation for a finite size number of x-ray reflecting planes, $I = A[\sin(Nqd_{QL})/\sin(qd_{QL})]^2 + B$, where $A$ is the scattering amplitude, $N$ is the number of scattering planes, $d_{QL}$ is the distance between scattering planes (inter-QL lattice spacing),[24] and $B$ is a background adjusted to a polynomial of at most order 2, yields the results shown in Fig. 3. In this analysis, the finite resolution of the instrument (0.006 Å$^{-1}$) was accounted by taking an autoconvolution of the data. Fractional number of layers and disorder were taken into account by calculating a Gaussian distribution of scattering layers of width $\sigma_N$ with different integer values of $N$. For the samples that showed the finite size interference fringes around the Bragg peak [e.g., the 4, 10, and 20 QL sample data in Figs. 3(e) - (g)], the data were very sensitive to the number of QLs. For samples thicker than 20 QL, the finite size fringes were difficult to observe and the results were not as accurate. One reason may have been that, in addition to the lack of fringes, the structural coherence along the growth direction may be limited to approximately 34 nm, perhaps as a result of stacking faults, although this hypothesis would need to be verified using other techniques. Interestingly, it appears that the lattice parameter along the growth direction decreased with increasing film thickness, approaching the bulk lattice constant of 9.540 Å for the thicker samples. This decrease in lattice constant with increasing thickness indicates that the thin samples were somewhat strained with respect to the bulk crystal structure.

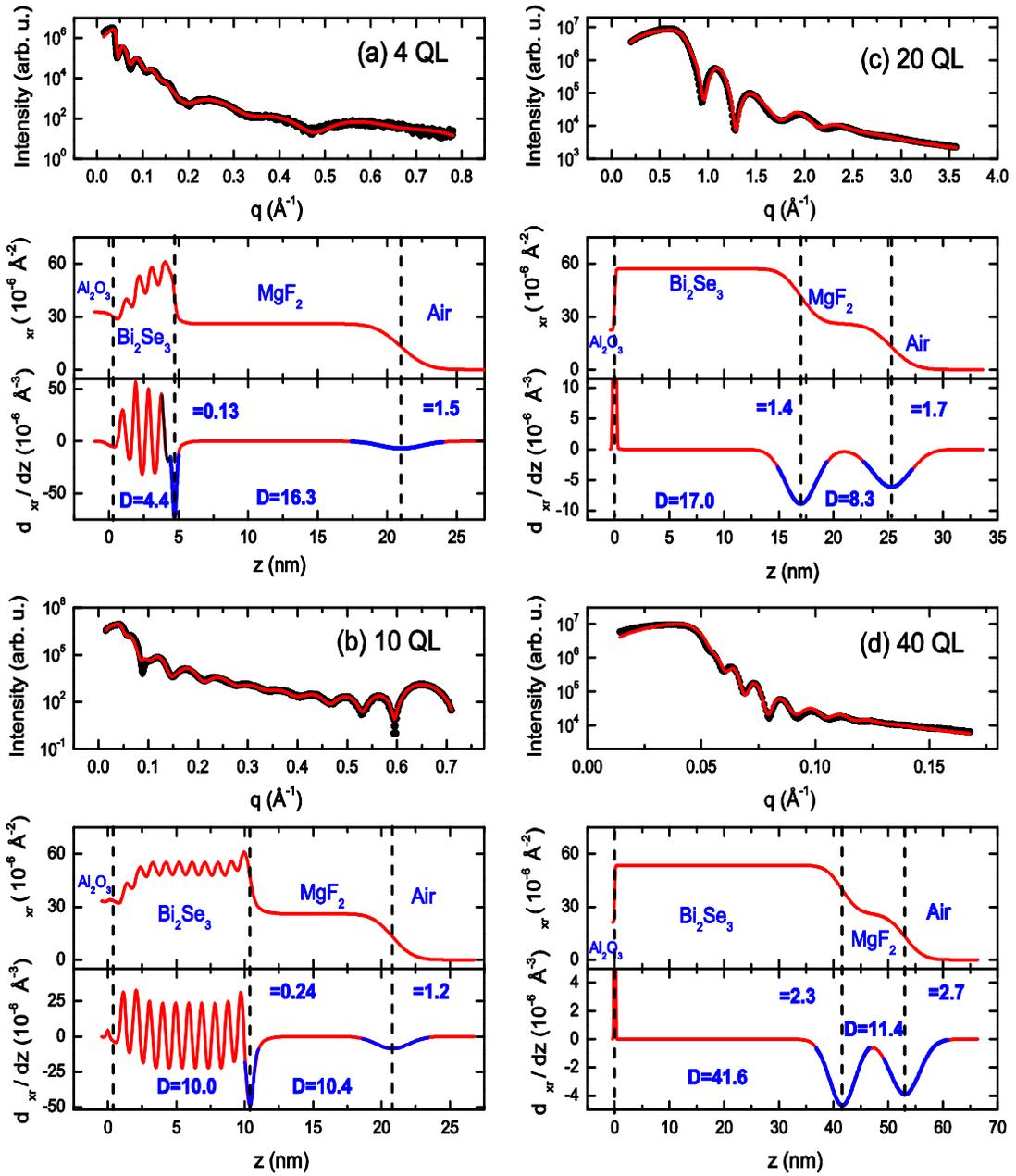

FIG. 2. X-ray reflectivity data (top row), the scattering length density $\rho_{xr}$ as a function of depth $z$ derived from the reflectivity (middle row), and the derivative of $\rho_{xr}$ as a function of $z$ (bottom row). The blue curves in the $d\rho_{xr}/dz$ graph indicate a fit to a Gaussian distribution whose center indicates the boundary between the layers and whose width is a measure of the roughness. The thickness of each layer $d$ and the roughness of the $Bi_2Se_3$/$MgF_2$ and $MgF_2$/air interfaces $\sigma$ are indicated in the figure. The roughness between substrate and $Bi_2Se_3$ was ~0.15 nm. The vertical dashed lines indicate the positions of the interfaces between the materials.

The XRD data show that the sample thicknesses determined using this methodology is consistent with the XRR

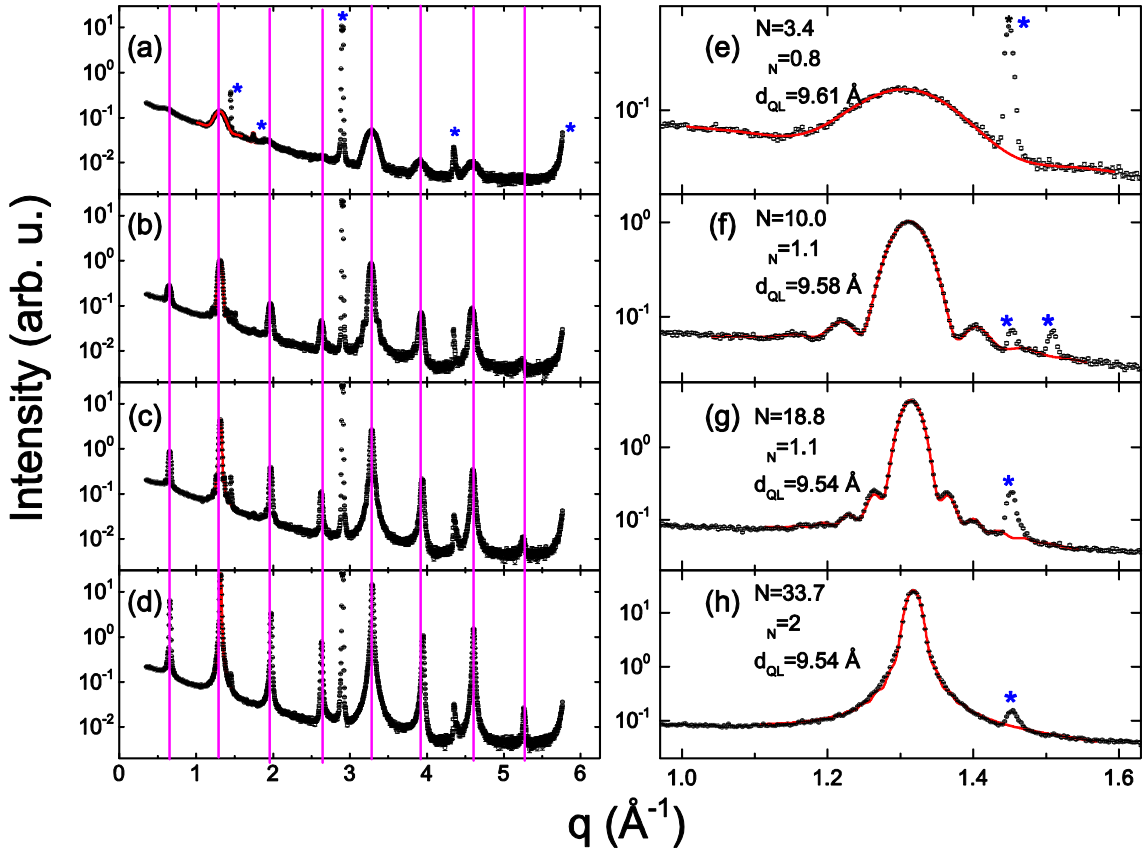

FIG. 3. X-ray diffraction data as a function of wavevector transfer perpendicular to sample surface q for the (a) 4 QL, (b) 10 QL, (c) 20 QL, and (d) 40 QL samples. The pink vertical lines indicate the position of the (003n) peaks (n is an integer), starting with n=1 peak on the far left. The panel on the right (e)-(h) show the data near the (006) peak along with the fits to the model described in the text (red curves). $N$, $\sigma_N$, and $d_{QL}$ are the average number of QLs, the disorder in the number of QLs, and the inter-QL lattice parameter used to fit the data, respectively. The peaks denoted by the blue asterisks are substrate or impurity peaks from the sample holder.

data. We conclude that as a whole, the samples are epitaxial and the nominal number of QLs is accurate to approximately 5%. The disorder increased with sample thickness, but all samples less than or equal 20 QL in thickness are of similar structural quality.

**B. Phonon dynamics determined from Raman scattering**

Figures 4(a) and (b) show Raman spectra measured from the top and bottom sides of the samples as indicated in the inset of Fig. 4(b). Comparing with our previous measurements,[9] the Raman spectra were re-measured for a broader range of $Bi_2Se_3$ film thicknesses from 2 to 40 nm (2 - 40 QLs) in order to cover the range where the direct intersurface coupling[25] and quantum confinement[26] effects (below 6 nm) may also contribute to the phonon dynamics. The spectra measured from the bottom side of the samples [Fig. 4(b)] are almost identical to that of the sapphire substrate.[9,27] Negligible variations of Raman intensities for the bottom side measured spectra suggest that the corresponding Raman intensity variations in the top side measured spectra is due to the $Bi_2Se_3$ film thickness effect on the Raman scattering efficiency. Because Raman spectra below 100 $cm^{-1}$ were significantly weakened due to the optical filter of the spectrometer used, we analyzed only four Raman features peaked at ~130 $cm^{-1}$ (3.9 THz), ~173 $cm^{-1}$ (5.2 THz), ~300 $cm^{-1}$ (9.0 THz), and ~252 $cm^{-1}$ (7.56 THz), which were assigned to the bulk $E_g^2$ (in-plane, in-phase), $A_{1g}^2$ (out-of-plane, out-of-phase), combined $E_g^2 + A_{1g}^2$ phonon modes associated with Bi-Se bonding[9,28-33] and the 2D surface phonon mode (H-mode) associated with Se-Se bonding of the topmost hexagonally arranged Se layer,[9] respectively. The variation of the $Bi_2Se_3$ film thickness also affects the positions of some of the Raman peaks and their linewidths. This behavior can be seen more clearly in Fig. 4(c) which shows the same Raman spectra presented in Fig. 4(a) but normalized at the peak position of the bulk $E_g^2$ phonon mode. It should be noted that in addition to the Raman features mentioned, there exists a broadband tail extending up to ~600 $cm^{-1}$, the intensity of which also varies with the film thickness. All these $Bi_2Se_3$ film thickness effects will be discussed further below.

Intensities of Raman peaks tend to increase with decreasing film thickness ($d$) in the range from ~40 nm to ~6 -

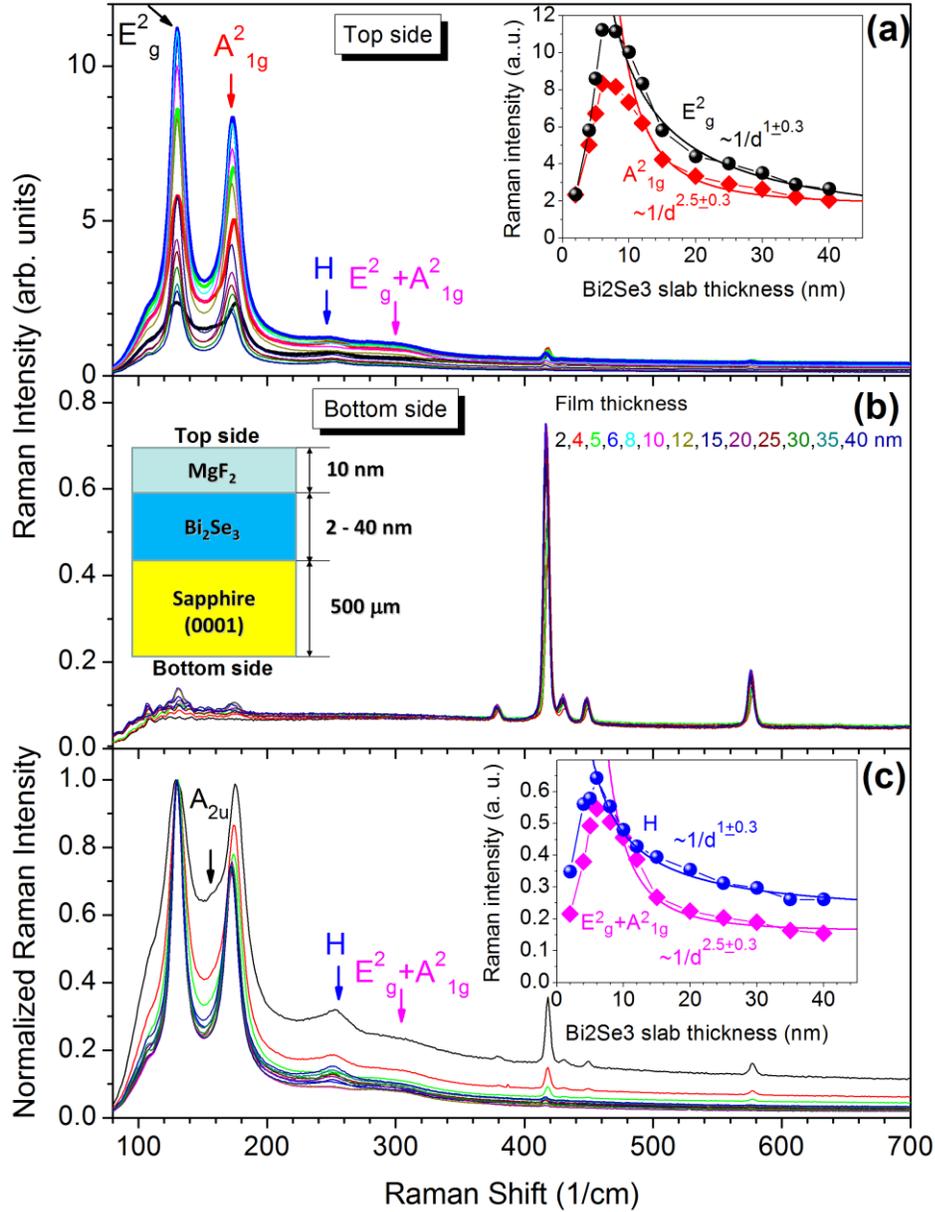

FIG. 4. Raman spectra measured from the top (a) and bottom side (b) of the samples [inset in (b)] of various thickness $Bi_2Se_3$ films as indicated by the corresponding colors. Spectra of the 2, 4, 5, and 6 nm thick films are presented in (a) using broader lines. (c) The same Raman spectra shown in (a) but normalized at the peak position of the bulk $E_g^2$ phonon mode. The increased background level in the normalized spectra of the 2, 4, and 5 nm thick films is due to the Raman peak intensity drop, although the background level for all of the samples is about the same. Insets in (a) and (c) show the film thickness dependences of Raman intensities of the $E_g^2$, $A_{1g}^2$ and H-mode, $E_g^2 + A_{1g}^2$ phonon modes, respectively.

8 nm, as $1/d^{1\pm0.3}$ for the in-plane phonon modes ($E_g^2$ and H-mode) and as $1/d^{2.5\pm0.3}$ for the out-of-plane phonon modes ($A_{1g}^2$ and $E_g^2 + A_{1g}^2$) [Fig. 4, Insets in (a) and (c)]. This behavior has been suggested to result from the resonant Raman enhancement due to nonlinear excitations of TM (acoustic) and TE (optical) plasmons, respectively.[9] A subsequent sharp decrease of Raman intensities with decreasing film thickness below ~6 - 8 nm can be associated with the incident laser light penetration length, which becomes longer than the film thickness,[6] thus significantly diminishing the Raman scattering efficiency due to a decrease in the scattering volume. Similar dynamics originating from the same mechanism have also been observed for graphene multilayers with decreasing the number of graphene layers.[34] Consequently, the Raman intensity dynamics in films whose thickness ranges from 2 to 8 nm is expected to be convoluted with the light penetration dynamics. To avoid this problem, we deconvolved Raman intensity dynamics with decreasing film thickness by dividing the Raman intensities by that of the $A_{1g}^2$ out-of-plane mode, which seems to be most sensitive to the light-penetration/film-thickness ratio. Figures 5(a) and (b) show the thickness dependences of the resulting deconvolved in such a manner (normalized) intensities ( $I_{E_g^2}/I_{A_{1g}^2}$ and $I_{E_g^2+A_{1g}^2}/I_{A_{1g}^2}$ ), which now take into account the effects of the light penetration depth in samples of different thicknesses. The normalized intensity of the $E_g^2$ in-plane phonon mode slightly increases with decreasing $d$ until $d$ ~6 nm and afterwards abruptly decreases for thinner films, whereas the normalized intensity of the combined $E_g^2+A_{1g}^2$ phonon mode shows the opposite dynamics. The gradual monotonic increase (decrease) of $I_{E_g^2}/I_{A_{1g}^2}$ ( $I_{E_g^2+A_{1g}^2}/I_{A_{1g}^2}$ ) normalized intensities with decreasing film thickness from 40 to 6 nm is due to the different trends of Raman intensity enhancement [Fig. 4, Insets in (a) and (c)]. The subsequent abrupt decrease (increase) of $I_{E_g^2}/I_{A_{1g}^2}$ ( $I_{E_g^2+A_{1g}^2}/I_{A_{1g}^2}$ ) normalized intensities for thinner films coincides with the opening of a gap in Dirac SS previously observed using the angle-resolved photoemission spectroscopy (ARPES)[25] and therefore it may be related to a crossover of the 3D TI Bi$_2$Se$_3$ to the 2D limit (gapped SS) due to the time-reversal symmetry breaking as a consequence of direct coupling between the opposite-surface Dirac SS. Because of the increase of the combined $E_g^2+A_{1g}^2$ phonon mode intensity for films with $d < 6$ nm, we associate this crossover in vibrational dynamics with anharmonic coupling between in-plane and out-of-plane bulk phonon modes induced by direct intersurface coupling in Dirac SS.

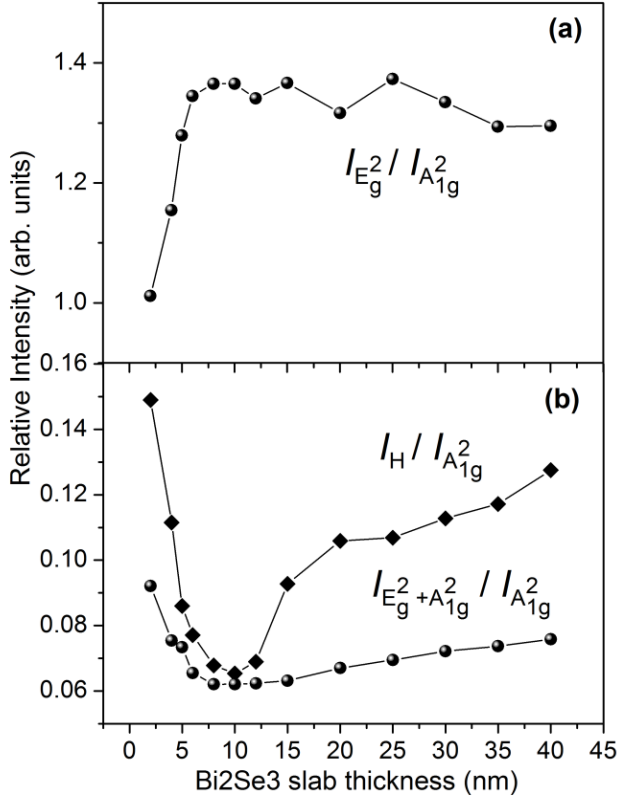
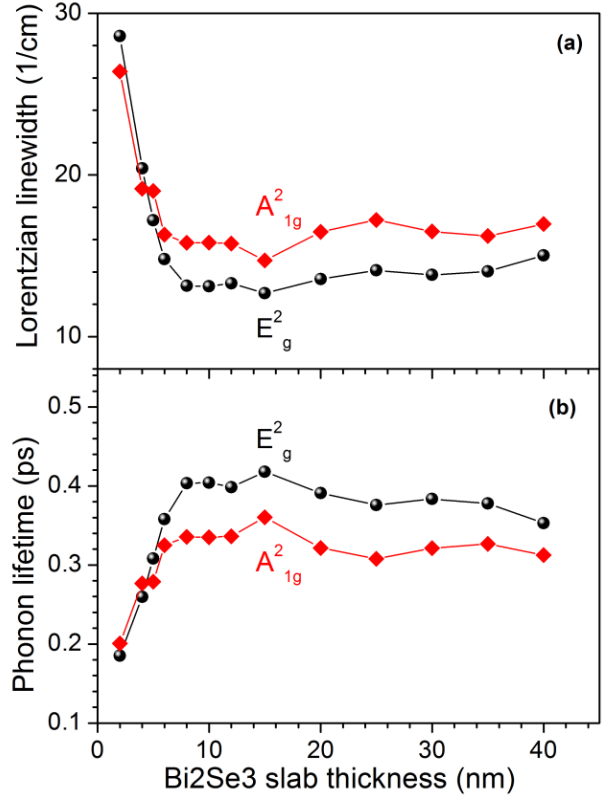

FIG. 5. Thickness dependences of the deconvolved (normalized at the peak position of the bulk $A_{1g}^2$ phonon mode) intensities of Raman peaks shown in Fig. 4 (see text).

FIG. 6. Thickness dependences of Lorentzian linewidths of the $E_g^2$ and $A_{1g}^2$ bulk phonon modes (a) and the corresponding phonon lifetimes (b).

Importantly, the $E_g^2$ and $A_{1g}^2$ phonon modes also reveal a substantial broadening of vibrational transitions with decreasing film thickness in the range below 6 nm [Fig. 4(c)]. The broadening dynamics can be recognized explicitly by fitting Raman peaks with Lorentzian profiles to obtain the thickness dependence of linewidth Γ (full width at half maximum). Figure 6(a) shows that the linewidth progressively increases exclusively for the film thickness range where the normalized intensity of the combined $E_g^2+A_{1g}^2$ phonon mode increases ($d < 6$ nm), thus suggesting the anharmonic nature of the broadening. Additionally, the infrared-active (Raman-inactive) $A_{2u}$ phonon mode peaked at ~159 cm$^{-1}$ weakly emerges for the thinnest film ($d = 2$ nm), indicating the crystal-symmetry breaking in ultrathin films [Fig. 4(c)].[35] However, because this Raman feature is extremely weak, the crystal-symmetry breaking is expected to be negligible. The latter fact additionally proves that the considerable broadening of the $E_g^2$ and $A_{1g}^2$ phonon mode peaks together with the increased intensity of the combined $E_g^2 + A_{1g}^2$ phonon mode peak can be associated with anharmonic coupling between in-plane and out-of-plane bulk phonon modes rather than with a possible crystalline structure imperfections of ultrathin films. Relating the phonon lifetime to the linewidth as $\tau_{ph} = \hbar/\Gamma$,[7] where $\hbar = 5.3\times10^{-12}$ cm$^{-1}$s is the reduced Planck's constant, one can recognize that the bulk phonon lifetime significantly decreases with decreasing film thickness below 6 nm [Fig. 6(b)]. This observation also points to strong anharmonic coupling between in-plane and out-of-plane bulk phonon modes, which can be an important reason for the reduced electron-phonon coupling strength[36] in the bulk of ultrathin films oppositely to the QL-to-QL stacking effect.[26]

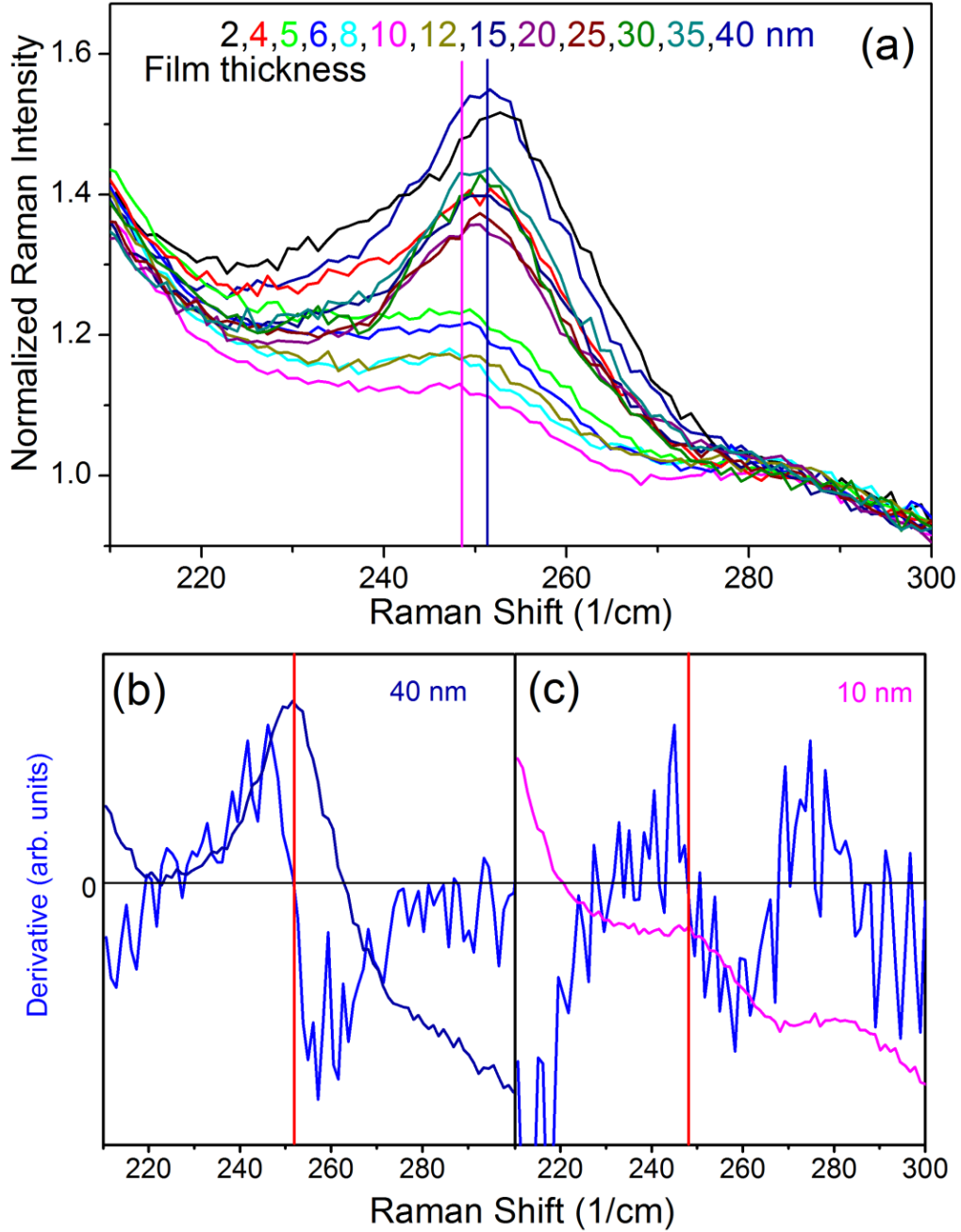

FIG. 7. (a) The H-mode range Raman spectra shown in Fig. 1(a) but normalized at 285 cm$^{-1}$. The corresponding color vertical line indicates the position of the peak for the 10 and 40 nm thick films. (b) and (c) The examples of the peak position determination of the H-mode for the 40 nm and 10 nm thick films. The peak positions were determined by differentiating the corresponding Raman features and estimating the Raman shift position (shown as the vertical red lines), at which the derivative is equal 0. The accuracy of this procedure is found to be below the step size of measurements (1 cm$^{-1}$).

The normalized Raman intensity of the surface H-mode ($I_H/I_{A_{1g}^2}$) shows non-monotonic behavior with decreasing film thickness with a minimum at ~10 nm [Fig. 5(b)]. These dynamics unambiguously distinguish between the bulk and surface phonon modes and points to another source of finite size effects on the Raman intensity dynamics that have been associated with the excitation of a Fano-type resonance due to strong plasmon-phonon coupling in the indirectly (electrostatically) coupled opposite-surface Dirac SS.[9] In this picture the plasmon modes are the collective electronic excitations of the coupled system, similar to what occurs in electrostatically coupled graphene bilayers.[37-41] The out-of-phase plasmon mode appearing at small wavevectors beyond the single particle continuum (the acoustic Dirac plasmon) that is responsible for the enhancement of the electron-phonon coupling strength in 2D Dirac SS is expected to be nonlinearly excited[9] within a long wavelength gap forming in the presence of electrostatic intersurface coupling, which allows for intersurface electron tunneling.[42] It should also be noted that the excitation of plasmons in $Bi_2Se_3$ thin films has been suggested to be responsible for the resonant enhancement of optical second-harmonic generation for the ~10 nm thick film.[43] The H-mode intensity decrease to a minimum for the ~10 nm thick film can hence be associated with quantum interference between the ionic-subsystem Raman response (surface H-mode) and the electronic-subsystem Raman response (the acoustic Dirac plasmon) that is shifted in phase. The rapid variation in phase gives rise to the asymmetric lineshape of the Fano-type resonance accompanied by the corresponding decrease of the relative peak intensity (destructive interference) when the electron-phonon coupling becomes stronger for a certain film thickness, as appeared for the thickness range of 5 – 12 nm (Fig. 7). Consequently, the lineshape of the H-mode Raman peak is close to Lorentzian for films thicker than 12 nm and thinner than 5 nm. However, the lineshape becomes a Fano-type (dispersion-like) for films with $d = 5 – 12$ nm. We note that the lineshape of Fano resonance varies similarly to that observed in infra-red absorption of bilayer graphene as a function of the bandgap energy tuned by applying a gate voltage or changing electron density.[44]

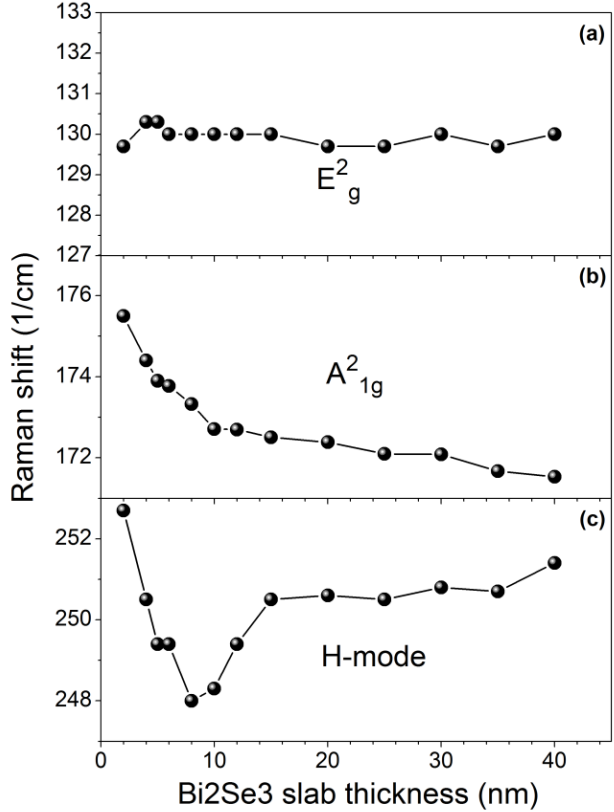

FIG. 8. Thickness dependences of the peak positions of the bulk ($E_g^2$ and $A_{1g}^2$) and surface (H-mode) phonon modes as indicated in (a), (b), and (c), respectively.

The shift of the bulk and surface phonon modes with decreasing film thickness additionally distinguishes between them. The in-plane phonon mode $E_g^2$ remains almost unshifted whereas the out-of-plane phonon mode $A_{1g}^2$ shows a significant monotonic shift toward higher frequencies (blue-shift) with decreasing film thickness [Figs. 8(a) and (b)]. We note that quantum confinement usually leads to a phonon softening (red-shift) of the bulk modes, thus eliminating this mechanism from consideration.[35] We associate the observed Raman peak shift trends for the in-plane and out-of-plane bulk phonon modes with the effect of indirect intersurface coupling on the atomic bonding unscreening dynamics due to long-range Coulomb interactions caused by the bulk carrier depletion.[7,9] Because indirect intersurface coupling acts predominantly along the film normal and therefore mainly affects the out-of-plane phonon mode, only the $A_{1g}^2$ mode shows a significant blue-shift (stiffening) with decreasing film thickness [Figs. 8(a) and (b)]. This behavior is well consistent with the transient stiffening of the G-mode phonons in graphite, which has been associated with a decrease in the electron-phonon coupling strength upon high electronic temperatures and the corresponding reduction in the phonon energy renormalization which depends on the net electron density.[36] On the other hand, the shift of the surface H-mode Raman peak reveals a non-monotonic behavior with a minimum at ~8 - 10 nm and in general follows the dynamics of the Raman intensity shown in Fig. 5(b). This coincidence suggests that indirect intersurface coupling and plasmon excitation effects are related to each other. The most important question to be addressed is related to plasmon-phonon coupling, which is expected to enhance the common electron-phonon coupling at higher electron densities satisfying the resonant frequency condition of $v_{pl} \approx v_H$,[9,45] where $v_{pl}$ and $v_H$ are the acoustic Dirac plasmon and surface phonon frequencies, respectively. Furthermore, because plasmon-phonon coupling is expected to take place if the surface H-mode (associated with Dirac SS) and the bulk in-plane phonon mode ($E_g^2$) remain uncoupled to the bulk out-of-plane phonon mode ($A_{1g}^2$), their anharmonic coupling with decreasing film thickness establishes a natural limit for plasmon-phonon coupling to occur. This behavior explains the non-monotonic thickness dependence of the H-mode Raman peak intensity and Raman shift since the carrier density in Dirac SS is known to increase with decreasing film thickness,[9] thus adjusting the acoustic Dirac plasmon frequency $v_{pl}$ to $v_H$ and hence increasing the electron-phonon coupling strength until the anharmonic coupling between in-plane and out-of-plane modes becomes important and starts to diminish it. Importantly, the enhancement of electron-phonon coupling in 2D Dirac SS is accompanied by the corresponding stiffening of the bulk $A_{1g}^2$ mode, thus indication the weakening of the electron-phonon coupling in the bulk [Figs. 8(b) and (c)]. We also note that the non-monotonic thickness dependence of the electron-phonon coupling strength with a maximum at 8 - 10 nm cannot be associated with the oscillatory behavior of gapped Dirac SS since these oscillations are predicted to appear for films thinner than 6 nm.[46-48] Furthermore, the anharmonic coupling between surface and bulk phonon modes may lead to an additional combined Raman peak at ~550 cm$^{-1}$. The aforementioned broadband tail in Raman spectra extending up to ~600 cm$^{-1}$ [Figs. 4(a) and (c)], the intensity of which varies with the film thickness close to the variation of intensities of the surface H-mode and the bulk $E_g^2$ and $A_{1g}^2$ modes, could result from this coupling. However, due to a small coupling strength and multicomponent nature, this Raman feature is weakly resolvable [Fig. 4(c)].

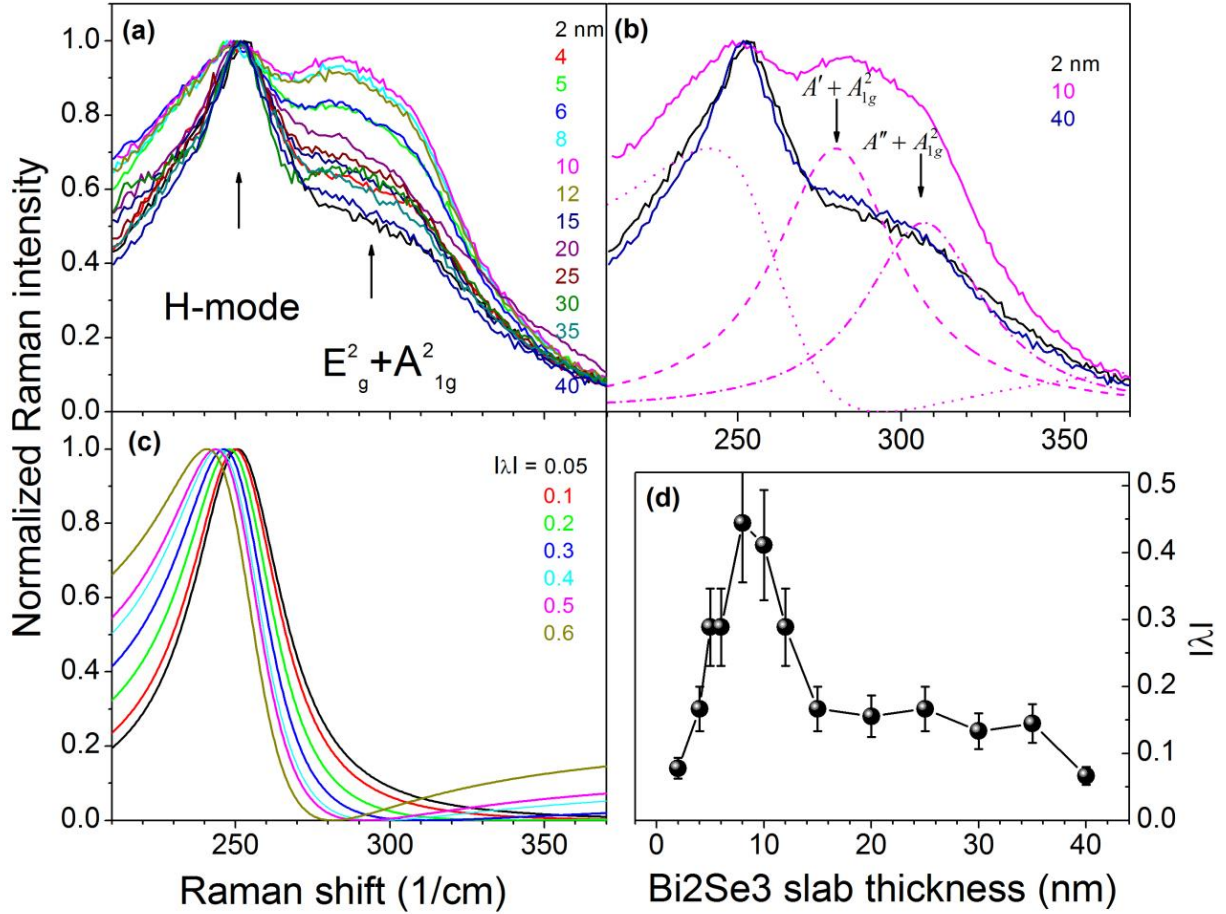

FIG. 9. (a) The normalized Raman spectra in the range of H-mode and $E_g^2 + A_{1g}^2$ phonon modes from which the $A_{1g}^2$ peak Lorentzian tail has been subtracted are shown for various $Bi_2Se_3$ film thicknesses as indicated by the corresponding colors. (b) Several spectra from those shown in (a) emphasize the non-monotonic dynamics of the broadening and shift of the H-mode Raman peak. Doted, dashed, and dash-dotted lines show an example of the fit by Lorentzian/Fano lineshapes for the 10 nm thick film. (c) Theoretical predictions for the H-mode broadening and shift are shown for various λ as indicated by the corresponding colors. (d) Electron-phonon coupling strength (λ) for various $Bi_2Se_3$ film thicknesses obtained from the H-mode shift.

The two-source Raman scattering process determining the appearance of the Fano-type resonance is typical for heavily doped semiconductors,[9,26,49,50] including $Bi_2Se_3$ ($n \sim 10^{19}$ cm$^{-3}$). The corresponding asymmetry parameter is known to characterize the electron-phonon coupling strength. Consequently, the H-mode Raman peak reveals the film-thickness dependent asymmetry that can be treated by the Fano resonance lineshape $I(\nu) = A[f + 2(\nu - \nu_H)/\Gamma_H]^2 / 1 + [2(\nu - \nu_H)/\Gamma_H]^2$,[9,25,26] where $A$ is a proportionality coefficient, $\nu_H$ and $\Gamma_H$ are the frequency and linewidth, respectively, of the surface H-mode peak without coupling to the electronic subsystem (of the Lorentzian lineshape $I(\nu) = A(\Gamma_H/2)^2 / [(\Gamma_H/2)^2 + (\nu - \nu_H)^2]$), and $f$ is the Fano asymmetry parameter related to the electron-phonon coupling strength constant as $\lambda = 1/f$. It is clear that if $\lambda \to 0$, the Fano resonance lineshape transforms to that of the Lorentzian lineshape.

To treat the Fano-type resonance in order to estimate the electron-phonon coupling strength, it is required to extract its intensity from those of the neighboring $A_{1g}^2$ and $E_g^2 + A_{1g}^2$ peaks. However, this kind of background subtraction seems to be inaccurate since there are no any criteria to identify the Fano lineshape, except for the Fano asymmetry parameter that in turn has to be determined. Figures 9(a) and (b) show an example of such a subtraction of the intensity of the $A_{1g}^2$ peak Lorentzian tail. The resulting normalized Raman spectra show that the H-mode peak asymmetrically broadens for film thicknesses ranging from 5 to 12 nm. A more precise analysis of the combined $E_g^2 + A_{1g}^2$ phonon mode as a function of the film thickness allowed us to recognize its splitting, which mainly manifests itself for films with $d$ ranging from 5 to 12 nm. Figure 9(b) shows that when the relative intensity of the $E_g^2 + A_{1g}^2$ mode increases, the corresponding peak splits into two components peaked at 285 and 308 nm. This splitting is suggested to result from the aforementioned anharmonic coupling between in-plane and out-of-plane modes, which leads to the degeneracy lifting of the $E_g^2$ phonon mode ($E_g^2 \to A' + A''$), thus allowing the two combined phonon modes ($A' + A_{1g}^2$ and $A'' + A_{1g}^2$) to appear in Raman spectra. The theoretical modeling of the Fano resonance [Fig. 9(c)] qualitatively depicts the dynamics of H-mode Raman peak with film thickness variations [Fig. 9(a)]. However, a quantitative treatment of the Fano resonance seems to be problematic since the asymmetry of extracted H-mode peak strongly depends on the background subtraction procedure used [Fig. 9(b)], which is far from unambiguous and therefore may lead to an underestimate of the electron-phonon coupling strength.[9]

On the other hand, the theoretical modeling shown in Fig. 9(c) proves that for weak electron-phonon coupling the Fano lineshape transforms to the Lorentzian lineshape, whereas the Fano resonance asymmetrically broadens and shifts toward the lower frequency range (red shift) with increasing the electron-phonon coupling strength. The resulting shift of the Fano resonance maximum ($\nu_{max}$) depends on the sign of $\lambda$, which for the measured H-mode Raman peak is negative, and on the absolute value of $\lambda$. The latter statement can be incorporated using the standard extremum-finding condition applying for the Fano resonance lineshape $dI(\nu)/d\nu = 0$, thus yielding $|\lambda| = 2|\nu_{max} - \nu_H|/\Gamma_H$. Consequently, we used the thickness dependence of the H-mode Raman peak shift shown in Fig. 8(c) and $\Gamma_H = 18$ cm$^{-1}$ for the relatively symmetric peak of the 40 nm thick film which for many measured parameters corresponds to bulk Bi$_2$Se$_3$.[6-9] Subsequently, the value $\nu_H = 252.1$ cm$^{-1}$ has been chosen to obtain for the 40 nm thick film $\lambda \sim 0.08$, which corresponds to the single Bi$_2$Se$_3$ crystals.[2]

This adjustment to the bulk value allowed us to demonstrate the finite size effect on the electron-phonon coupling strength in Dirac SS [Fig. 9(d)]. For thicker than 12 nm films (15 - 35 nm) and for thinner than 5 nm films (2 - 4 nm), the electron-phonon coupling strength corresponds to the bulk value λ ~0.1. However, the electron-phonon coupling is about four-fold plasmon-enhanced with respect to the bulk value when the film thickness ranges from 8 to 10 nm. The maximal value of λ = 0.44 obtained for the 8 nm thick film approaches the value of λ = 0.62 for Cu$_x$Bi$_2$Se$_3$ and Bi$_2$Te$_3$,[2] for which the superconductivity upon Cu-doping[51,52] and under pressure[53] has been reported, respectively. We also note that the plasmon-enhanced electron-phonon coupling in Dirac SS for film thicknesses ranging from 8 to 10 nm is comparable to the bulk electron-phonon coupling (~0.43) deduced from inelastic helium-atom scattering,[1] which can be associated with electron-polar-phonon (Fröhlich) interaction.[6-8] This coincidence can be a reason for confusing between estimations of the electron-phonon coupling strength in the bulk and 2D Dirac SS of Bi$_2$Se$_3$ films. We note also that this resonance-type enhancement of the electron-phonon coupling in Dirac SS of the ~8 - 10 nm thick Bi$_2$Se$_3$ film, which is accompanied by the corresponding weakening of the electron-phonon coupling in the bulk, has been observed directly using an ultrafast transient reflectivity technique as a resonant decrease of the carrier relaxation rate, which mainly reflects the bulk carrier dynamics.[6] The nature of this effect becomes clear only now.

## IV. CONCLUSIONS

Epitaxial samples of Bi$_2$Se$_3$ with nominal thicknesses ranging from 2 to 40 QL were grown on Al$_2$O$_3$ substrate and their phonon dynamics were analyzed using Raman scattering. Our findings indicate that the electron-phonon coupling in Dirac SS of thin-films of the TI Bi$_2$Se$_3$ can be significantly enhanced due to plasmon excitation at higher electron densities satisfying the resonant condition, at which the plasmon frequency matches the surface phonon frequency. Because the free electron density in Dirac SS is known to increase with decreasing film thickness, the plasmon frequency resonance ($\nu_{pl} \approx \nu_H$) enhances the electron-phonon coupling strength at a certain film thickness until the anharmonic coupling between in-plane and out-of-plane modes diminishes it for thinnest films. This competition between the two processes governs a non-monotonic character of the electron-phonon coupling strength in 2D Dirac SS with a maximum for the 8 - 10 nm thick films. The observed plasmon-enhanced electron-phonon coupling in Dirac SS about four-fold exceeds that known for the bulk Bi$_2$Se$_3$. We concluded that this observation may provide new insights into the origin of superconductivity in this-type materials in particular and layered structures in general.

**ACKNOWLEDGMENTS**

We thank T. A. Johnson for helping with the growth of the MgF$_2$ capping layer in some of the original samples. This work was supported by a Research Challenge Grant from the West Virginia Higher Education Policy Commission (HEPC.dsr.12.29). Some of the work was performed using the West Virginia University Shared Research Facilities.

[1] X. Zhu, L. Santos, C. Howard, R. Sankar, F. C. Chou, C. Chamon, and M. El-Batanouny, Phys. Rev. Lett. **108**, 185501 (2012).
[2] Z.-H. Pan, A.V. Fedorov, D. Gardner, Y. S. Lee, S. Chu, and T. Valla, Phys. Rev. Lett. **108**, 187001 (2012).
[3] J. A. Sobota, S.-L. Yang, D. Leuenberger, A. F. Kemper, J. G. Analytis, I. R. Fisher, P. S. Kirchmann, T. P. Devereaux, and Z.-X. Shen, Phys. Rev. Lett. **113**, 157401 (2014).
[4] S. Murakami, New J. Phys. **9**, 356 (2007).
[5] M. Z. Hasan, C. L. Kane, Rev. Mod. Phys., **82**, 3045 (2010).
[6] Y. D. Glinka, S. Babakiray, T. A. Johnson, A. D. Bristow, M. B. Holcomb, and D. Lederman, Appl. Phys. Lett. **103**, 151903 (2013).
[7] Y. D. Glinka, S. Babakiray, T. A. Johnson, M. B. Holcomb, and D. Lederman, Appl. Phys. Lett. **105**, 171905 (2014).
[8] Y. D. Glinka, S. Babakiray, T. A. Johnson, M. B. Holcomb, and D. Lederman, J. Appl. Phys. **117**, 165703 (2015).
[9] Y. D. Glinka, S. Babakiray, T. A. Johnson, and D. Lederman, J. Phys.: Condens. Matter **27**, 052203 (2015).
[10] P. M. R. Brydon, S. Das Sarma, H.-Y. Hui, and J. D. Sau, Phys. Rev. B **90**, 184512 (2014).
[11] X. Wan and S. Y. Savrasov, Nature Commun. **5**, 4144 (2014).
[12] X.-L. Zhang and W.-M. Liu, Sci. Rep. **5**, 8964 (2015).
[13] I. Loa, E. I. Isaev, M. I. McMahon, D. Y. Kim, B. Johansson, A. Bosak, and M. Krisch, Phys. Rev. Lett. **108**, 045502 (2012).
[14] Z. P. Yin, S. Y. Savrasov, and W. E. Pickett, Phys. Rev. B 74, 094519 (2006).
[15] R. E. Cohen, W. E. Pickett, and H. Krakauer, Phys. Rev. Lett. 64, 2575 (1990).
[16] F. Weber, S. Rosenkranz, L. Pintschovius, J.-P. Castellan, R. Osborn, W. Reichardt, R. Heid, K.-P. Bohnen, E. A. Goremychkin, A. Kreyssig, K. Hradil, and D. L. Abernathy, Phys. Rev. Lett. **109**, 057001 (2012).
[17] P. Zhang, S. G. Louie, and M. L. Cohen, Phys. Rev. Lett. **98**, 067005 (2007).
[18] K. Kirshenbaum, P. S. Syers, A. P. Hope, N. P. Butch, J. R. Jeffries, S. T. Weir, J. J. Hamlin, M. B. Maple, Y. K. Vohra, and J. Paglione, Phys. Rev. Lett. **111**, 087001 (2013).
[19] S. Mandal, R. E. Cohen, and K. Haule, Phys. Rev. B **89**, 220502 (2014).
[20] P. Tabor, C. Keenan, S. Urazhdin, and D. Lederman, Appl. Phys. Lett. **99**, 013111 (2011).
[21] M. Björck and G. Andersson, J. Appl. Cryst. **40**, 1174 (2007).
[22] K. Munbodh, F. A. Perez, C. Keenan, D. Lederman, M. Zhernenkov, and M. R. Fitzsimmons, Phys. Rev. B **83**, 094432 (2011).
[23] P. Lostak, C. Drasar, H. Sussmann, P. Reinshaus, R. Novotny, and L. Benes, J. Cryst. Growth **170**, 144 (1997).
[24] B. E. Warren, *X-Ray Diffraction* (Dover, New York, 1990).
[25] Y. Zhang, K. He, C.-Z. Chang, C.-L. Song, L.-L. Wang, X. Chen, J.-F. Jia, Z. Fang, X. Dai, W.-Y. Shan, S.-Q. Shen, Q. Niu, X.-L. Qi, S.-C. Zhang, X.-C. Ma, and Q.-K. Xue, Nature Phys. **6**, 584 (2010).
[26] L. He, F. Xiu, X. Yu, M. Teague, W. Jiang, Y. Fan, X. Kou, M. Lang, Y. Wang, G. Huang, N.-C. Yeh, and K. L. Wang, Nano Lett. **12**, 1486 (2012).
[27] M. Kableikova, J. Breza, and M. Vesely, Microelect. J. **32**, 955 (2001).
[28] J. Zhang, Z. Peng, A. Soni, Y. Zhao, Y. Xiong, B. Peng, J. Wang, M. S. Dresselhaus, and Q. Xiong, Nano Lett. **11**, 2407 (2011).
[29] V. Gnezdilov, Yu. G. Pashkevich, H. Berger, E. Pomjakushina, K. Conder, and P. Lemmens, Phys. Rev. B **84**, 195118 (2011).
[30] I. W. Richter, H. Kohler, and C. R. Becke, Phys. stat. sol. **84**, 619 (1977).
[31] I. Childres, J. Tian, I. Miotkowski, Y. Chen, Phil. Mag. **93**, 681 (2012).
[32] K. M. F. Shahil, M. Z. Hossain, V. Goyal, and A. A. Balandin, J. Appl. Phys. **111**, 054305 (2012).
[33] X. X. Yang, Z. F. Zhou, Y. Wang, R. Jiang, W. T. Zheng, and C. Q. Sun, J. Appl. Phys. **112**, 083508 (2012).
[34] P. Klar, E. Lidorikis, A. Eckmann, I. A. Verzhbitskiy, A. C. Ferrari, and C. Casiraghi, Phys. Rev. B **87**, 205435 (2013).
[35] M. Eddrief, P. Atkinson, V. Etgens, and B. Jusserand, Nanotech. **25**, 245701 (2014).
[36] H. Yan, D. Song, K. F. Mak, I. Chatzakis, J. Maultzsch, and T. F. Heinz, Phys. Rev. B **80**, 121403 (2009).
[37] T. Stauber and G. Gomez-Santos, Phys. Rev. B **85**, 075410 (2012).
[38] T. Stauber, J. Phys.: Condens. Matter **26**, 123201 (2014).
[39] E. H. Hwang and S. Das Sarma, Phys. Rev. B **75**, 205418 (2007).
[40] E. H. Hwang and S. Das Sarma, Phys. Rev. B **80**, 205405 (2009).
[41] M. Jablan, H. Buljan, and M. Soljaci, Opt. Exp. **19**, 11236 (2011).
[42] S. Das Sarma and E. H. Hwang, Phys. Rev. Lett. **81**, 4216 (1998).
[43] Y. D. Glinka, S. Babakiray, T. A. Johnson, M. B. Holcomb, and D. Lederman, Phys. Rev. B **91**, 195307 (2015).
[44] T. T. Tang, Y. Zhang, C.-H. Park, B. Geng, C. Girit, Z. Hao, M. C. Martin, A. Zettl, M. F. Crommie, S. G. Louie, Y. R. Shen, and F. Wang, Nature Nanotechnol. **5**, 32 (2010).
[45] E. H. Hwang, R. Sensarma, and S. Das Sarma, Phys. Rev. B **82**, 195406 (2010).
[46] J. Linder, T. Yokoyama, and A. Sudbø, Phys. Rev. B **80**, 205401 (2009).
[47] C. X. Liu, H. J. Zhang, B. Yan, X. L. Qi, T. Frauenheim, X. Dai, Z. Fang, and S. C. Zhang, Phys. Rev. B **81**, 041307(R) (2010).
[48] S. Giraud, A. Kundu, and R. Egger, Phys. Rev. B **85**, 035441 (2012).
[49] J. F. Scott, T. C. Damen, J. Ruvalds, and A. Zawadowskit, Phys. Rev. B **3**, 1295 (1971).
[50] R. Kumar and A. K. Shukla, Phys. Lett. A **373**, 2882 (2009).
[51] Y. S. Hor, A. J. Williams, J. G. Checkelsky, P. Roushan, J. Seo, Q. Xu, H. W. Zandbergen, A. Yazdani, N. P. Ong, and R. J. Cava, Phys. Rev. Lett. **104**, 057001 (2010).

[52] M. Kriener, K. Segawa, Z. Ren, S. Sasaki, and Y. Ando, Phys. Rev. Lett. **106**, 127004 (2011).

[53] J. L. Zhang, S. J. Zhang, H. M. Weng, W. Zhang, L. X. Yang, Q. Q. Liu, S. M. Feng, X. C. Wang, R. C. Yu, L. Z. Cao et al., Proc. Natl. Acad. Sci. U.S.A. **108**, 24 (2010).